# Flows and Instabilities of Ferrofluids at the Microscale

Arthur Zakinyan, Elena Beketova, and Yuri Dikansky

Department of General and Theoretical Physics, Institute of Mathematics and Natural Sciences,

North Caucasus Federal University, 1 Pushkin Street, 355009 Stavropol, Russia

Correspondence should be addressed to Arthur Zakinyan; zakinyan.a.r@mail.ru

**Abstract** In the present work we report on the behavior of ferrofluid microdrops in an immiscible nonmagnetic carrier fluid and vice versa subjected to the action of magnetic fields. Our experiments evidence previously unexplored instability and flow patterns. We investigated the instabilities initiated by constant magnetic field and by combined action of constant and rotating magnetic fields. The observed instabilities lead to the formation of various microdrop configurations such as star-like and comb-like shapes, bending deformation, ridges and crests, spiral arms and rings, etc. The reviled peculiarities of the microscopic drops behavior can be of interest in applications to the ferrofluid-based microfluidics.

## 1 Introduction

Instability and flow patterns are exhibited in diverse physical, biological, and technological systems. It has been thought that study of the instability patterns of particular systems can provide insight into the mechanisms leading to the formation of similar appearing patterns. The ferrofluids are colloidal suspensions of ultra-fine ferromagnetic nanoparticles suspended in a carrier fluid (Blums et al. 1997). In general hydrodynamical problems the ferrofluid can be considered as continuous liquid magnetizable medium. Manipulated by an external magnetic field, the ferrofluids may demonstrate the wide variety of hydrodynamical instabilities (Rinaldi et al. 2005; Torres-Diaz and Rinaldi 2014). In this context, the ferrofluid can be regarded as an essential tool for the fluid flow and instability investigation (Timonen at al. 2013; Jamin et al.

2011). Previously a large number of various experimental and theoretical studies had been carried out regarding the behavior of ferrofluid macroscopic drops (of the order of 1 cm) under the external magnetic field. The stationary shape of ferrofluid drop under the action of constant magnetic field has been studied by Drozdova et al. (1981), Bacri and Salin (1982), Afkhami et al (2010), Zhu et al. (2011), Ody et al. (2016). The drop dynamics under the action of uniform rotating magnetic field has been considered by Lebedev et al. (2003). The instability of ferrofluid drops confined in a Hele-Shaw cell under the action of perpendicular magnetic fields has been investigated by Tsebers and Maiorov (1980), Dickstein et al. (1993), Elias et al. (1997), Bacri et al. (1988), Horng et al. (2001), Jackson and Miranda (2007), Chen et al. (2008), Chen et al (2009), Chen et al. (2010), Wen et al. (2011). The instability of flattened ferrofluid drops confined in a plane layer under the simultaneous action of in-plane rotating and normal magnetic fields has been considered by Elborai et al. (2005). Many fewer studies have been devoted to the investigation of inverse systems of nonmagnetic drops immersed in a ferrofluid. In this context the works of Tsebers and Blūms (1988) and Tatulchenkov and Cebers (2005) can be mentioned where the deformation of flattened bubble in a ferrofluid layer under the action of stationary normal magnetic field has been studied.

Recently potential new uses of the ferrofluid have arisen in the areas of development of ferrofluid-based microfluidic devices (Nguyen et al. 2007; Liu et al. 2011; Nguyen 2012) and controllable composite materials such as ferrofluid emulsions (Zakinyan and Dikansky 2011, 2017). It is important to know the behavior of ferrofluid microscopic drops in various situations as various surface tension- and hydrodynamic-related behaviors play important role in ferrofluid microdrops applications to the mentioned issues. However, thus far very few experimental or theoretical papers have addressed this promising new area. The works of Bacri et al. (1994), Janiaud et al. (2000), and Sandre et al. (1999) could be cited in this regard where the dynamics of ferrofluid microscopic droplets under the action of rotating magnetic field has been analyzed. As to investigation of nonmagnetic microdrops immersed in a ferrofluid, it seems that only the works of Dikansky and Zakinyan (2010) and Zakinyan et al. (2012) can be mentioned. In these works the nonmagnetic microdrop dynamics under the action of rotating magnetic field and combined action of electric and magnetic fields has been studied.

In view of the foregoing, the present research has two main goals. The first is to reveal the peculiarities of the behavior of ferrofluid drops arising at the transition from macro to micro level. The second is to investigate previously unexplored behavior of nonmagnetic microdrops immersed in a ferrofluid at the analogous conditions. In the present work we demonstrate some new futures of the drops behavior at the microscopic level arisen from competitions between the magnetic normal stress and the surface tension, the tangential stress and the drop viscous torque. Here we report on the experimental investigation of the behavior of ferrofluid microscopic drops in an immiscible nonmagnetic carrier fluid and vice versa in plane layer geometry under the simultaneous action of in-plane rotating and perpendicular static magnetic fields.

## 2 Experimental setup

In our experiments we used a kerosene-based ferrofluid with dispersed magnetite nanoparticles of about 10 nm diameter stabilized with oleic acid. The measured properties of the ferrofluid are: density is 1410 kg/m$^3$, dynamic viscosity is 21 mPa s, magnetite volume fraction is 14.5% and initial magnetic susceptibility is 3. The ferrofluid used was made by OJSC "NIPIgaspererabotka" (Russia). FH 51 aviation oil immiscible with the ferrofluid was selected as the nonmagnetic liquid. Its density is 840 kg/m$^3$ and the dynamic viscosity is 15 mPa s. The main reason to use this oil is that the interfacial tension at the interface between it and the ferrofluid is quite low ($\approx 10^{-6}$ N/m). The interfacial tension was determined by the retraction of the deformed drop method (Yu et al. 2004). The experimental sample was prepared by mechanical dispersing either a small volume of ferrofluid in a nonmagnetic liquid or a small volume of nonmagnetic liquid in ferrofluid. No additional stabilizing agents were used in the preparation process. The radii of microdrops obtained in this way are varying from 5 to 50 μm in zero fields. Since the size of magnetic nanoparticles of ferrofluid is much smaller than the sizes of microdrops, the ferrofluid can be considered as a continuous liquid magnetizable medium. The volume fraction of the suspended microdrops was about 0.01. Owing to the small volume fraction we could explore the behavior of a single drop and neglect the effects of drops interaction. Because of low interfacial tension, the microdrops are easily deformable in weak uniform magnetic fields on the order of a few 100 A/m.

The sample cell was assembled of two rectangular horizontal flat glass plates. A fluoroplastic film 40 μm thick with a circular hole in the center was placed between the plates to set the distance between them. The hole was filled with the studied sample. Some large microdrops were compressed by the plates. The microdrops were observed from above with an optical microscope and their behavior was recorded through a video camera. Alternating current is supplied with a phase shift of $\pi/2$ into two pairs of perpendicular Helmholtz coils to produce a uniform counterclockwise rotating magnetic field in the plane of the sample cell (Figure 1). The area of magnetic field uniformity was about 2 cm in diameter. The rotating field strength is ranged from 0 to 8 kA/m and the frequency is ranged from 5 to 1000 Hz. Another pair of coils is used to produce a uniform constant magnetic field orientated normally to the sample layer. The coils have been calibrated, and it has been measured that the current of 1 A produces the magnetic field of 1.9 kA/m at the area of homogeneity. It was possible to control the magnetic field establishment time in our experiments. Further we will distinguish two main regimes of magnetic field application: gradually rising magnetic field and instantaneously applied magnetic field. In the first case the magnetic field has been established by slowly increasing it from zero to some value (~ 50 A/(m·s)). In this case it can be considered that the drops have the stationary shapes at any time moment. In the second case the magnetic field of some value has been switched on within less than 0.01 s.

# 3 Microdrops confined in a plane layer under the action of perpendicular magnetic field

When ferrofluid drop is confined with an immiscible fluid between horizontal, closely spaced, parallel plates, and a uniform magnetic field is applied normal to the plates, the interface between fluids exhibits a threshold instability in which a comb-like pattern is established (Dickstein et al. 1993; Elias et al. 1997). The final state of this process is the formation of a labyrinth pattern. The ferrofluid drops breakup and ordered patterns in such systems have also been reported (Chen et al. 2008; Chen et al. 2010; Wen et al. 2011). We have found that the droplet behavior changes at the microscale when the interfacial tension is quite low. We studied the ferrofluid microdrop immersed in a horizontal flat layer subjected to a uniform gradually rising perpendicular magnetic field. With increasing the field strength, the drop is stretched along the direction of the field becoming almost cylindrical. When the magnetic field strength exceeds a critical value, the circular symmetry of the drop breaks and the sharp peaks of ferrofluid branch out of the drop at the contact with the upper and lower plates. The drop becomes star-shaped at the contacts with the plates (Figures 2a and 3a). The patterns formed at the upper and lower plates were observed by sequential focusing the microscope at these points. The number of peaks is proportional to the field strength, to the drop size, and to the field ramp rate. Thus the number of peaks is much larger at the instantaneously applied magnetic field in comparison with the slowly increasing magnetic field to the same strength.

If the magnetic field is further raised, the drop breakups into several smaller star-like drops (Figure 3). This breakup should be analogous in nature to that observed in previous studies (Horng et al. 2001) but it demonstrates a distinct pattern forming during the breakup. The instability behavior of the ferrofluid microdrop changes if the sufficiently strong magnetic field (higher than the breakup field) is turned on instantaneously. In this case the intensive droplet breakup into a large number of smaller droplets is observed. The instability development shows itself in the induced in-plane radially outward ferrofluid flow along several twisted trajectories (see Figure 2b and supplementary Online Resource 1). The number of trajectories is proportional to the field strength and to the drop size. Such behavior was not observed previously for the macroscopic drops.

If the layer contains sufficiently large number of closely spaced ferrofluid microdrops, the long range repulsion due to the interaction of parallel dipolar moments prevents the star-like shape formation and the drop breakup and leads to the labyrinth pattern formation (Figure 2d). In this case the instability behavior of the microdrops changes under the influence of neighboring microdrops; instead of branching into star-like shapes the drops undergo the bending instability. This bending instability is analogous to that observed previously (e.g., Dickstein et al. (1993), Jackson et al. (1994)) for a single macroscopic drops, but here, contrary to the previous results, such behavior takes place as a collective effect. The instability threshold for the array of closely spaced droplets has been found higher than that of individual microdrop. The blocking of the microdrop breakup by neighboring droplets also has not been observed previously.

The phenomena complementary to the above described are known when a drop of nonmagnetic fluid surrounded by ferrofluid and confined in a plane layer is stretched into a labyrinthine form due to application of a perpendicular magnetic field (Tsebers and Blūms 1988; Tatulchenkov and Cebers 2005). Contrary to the above experiments, the microscopic nonmagnetic drop instability pattern does not demonstrate any qualitative differences with the corresponding macroscopic case. At supercritical strengths of the magnetic field a mutual invasion of the two fluids takes place, in which fingers of ferrofluid invade the nonmagnetic fluid and vice versa. The developed viscous fingering pattern appears as the invasion process continues (see Figure 2c and supplementary Online Resource 1). Higher magnetic fields lead to thinner fingers, and higher field ramp rates lead to more branching. It was found that the critical magnetic field strength at which the circular symmetry of the microdrop becomes broken decreases with increasing microdrop size for the ferrofluid droplets and for the nonmagnetic droplets (Figure 4). This result correlates with the known results for macroscopic drops (Blums et al. 1997; Tsebers and Blūms 1988).

Next we consider the dynamics of the ferrofluid microscopic stripe confined in a horizontal layer under the instantaneously applied normal magnetic field. The stripe was prepared by applying the strong constant uniform in-plane magnetic field to a system of droplets. The adjacent droplets coalesce and stretch along the field producing the stripe. Owing to the low interfacial tension the stripe can hold its shape for a comparatively long time after the magnetic field switching off. Note that bending and alternated fingering instabilities of a ferrofluid macroscopic stripe have been studied previously (Bacri et al. 1995). Under the action of supercritical magnetic field we observed the appearance of sharp cones on the both sides of ferrofluid microstripe. This differentiates the case under study from the previous results where the smoothed fingers have been observed. Further evolution of the microstripe shape leads to bending into the sinusoidal geometry followed by microstripe breakup (Figure 2e). Thus here we have two successive instability types: at first the conical peaks are formed on the stripe and then (at higher fields) the bending of the stripe takes place. The simultaneous development of instabilities of two types on ferrofluid microstripes is another difference of the system under study from previously investigated macroscopic cases. Experiments with nonmagnetic fluid microstripe surrounded by ferrofluid did not shown qualitative differences with the macroscopic ferrofluid stripe behavior. The bending instability (Figure 2f) occurs for relatively thin microstripes and for low magnetic fields. The fingering instability (Figure 2g) occurs for relatively thick microstripes and for high magnetic fields (see Figure 5a). The instability wavelength decreases with the magnetic field and with the stripe thickness (see Figure 5b).

**4 Microdrops dynamics under the combined action of uniform rotating and perpendicular stationary magnetic fields**

The behavior of ferrofluid microdrops immersed in a horizontal flat layer under the action of in-plane-rotating magnetic field has been studied previously (Bacri et al. 1994; Janiaud et al. 2000; Sandre et al. 1999). It was shown that the microdrop can take a prolate ellipsoidal shape at low field frequencies or an oblate ellipsoidal shape at high field frequencies. The microdrops co-rotate with the field. S-like and starfish-like shapes can also be observed under certain system parameters. Simultaneously acting in-plane-rotating and stationary uniform normal magnetic fields can lead to the formation of spiral structures from the ferrofluid macroscopic drops confined in a plane layer (Elborai et al. 2005). As it will be shown below, the flows and instabilities of ferrofluid microdrops under the combined action of rotating and stationary magnetic fields may be much more complicated and varied. In our experiments at first the uniform in-plane-rotating field was applied and the static perpendicular field was superimposed. The field frequency was sufficiently high (always higher than the inverse characteristic relaxation time of a drop shape $\tau = 2\pi R(16\eta_e + 19\eta_i)(3\eta_e + 2\eta_i)/[40\sigma(\eta_e + \eta_i)]$, where $R$ is the initial radius of the droplet; $\sigma$ the interfacial tension at fluid-fluid interface; $\eta_{i,e}$ the dynamic viscosities of internal and external phases correspondingly), so the droplet was flattened in the horizontal plane and disk-like shaped. The formation of starfish shapes was prevented due to the high enough rotating field amplitude and frequency. If the stationary normal magnetic field is subsequently increased above a threshold value, the instability development is taking place. The instability character depends on the magnetic field strength and frequency.

At low frequencies and gradually rising perpendicular magnetic field the first stage of boundary deformation produces liquid ridges on the drop edge: a flat disk, crowned with ridges, that rotates as the field does (see Figure 6a and supplementary Online Resource 2). If the normal magnetic field is further raised, these ridges become unstable leading to the chaotically distorted drop boundary (Figure 6b). With the further increase in the normal field strength, the small rotating ferrofluid droplets begin to break away from the initial drop (Figure 6c). The final stage of instability development is the formation of large number of small ferrofluid droplets at high normal field strength. This array of droplets rotates as a whole with very low frequency (tenths or hundredths of a hertz). Analogous or any similar droplet shape evolution has not been observed in previous studies. At higher frequencies of rotating magnetic field the action of normal magnetic field leads to the formation of spiral arms of a microdrop (see Figure 6e and supplementary Online Resource 2). The formation of spiral arms has been known for macroscopic drops (Elborai et al. 2005) but, contrary to the previous result, in the current situation the arms are unstable: the development of arms results in the drop breakup into several rotating rings and individual small droplets (Figure 6f). At an even greater frequency the spiral arms become stable. The arms curl in on themselves, eventually forming the smooth rotating spiral pattern (Figure 6g).

The instability pattern also depends on the rotating field amplitude. Thus, at comparatively high amplitudes, the transformation of chaotically distorted boundary into the large crests was observed (Figure 6d). Such boundary instability pattern has not been observed in the previous studies. The further development of the instability patterns remain nearly the same as at low field amplitudes. The phase diagram of observed instability patterns is presented in Figure 7. It was found that the drop behavior depends on the normal field ramp rate. Consider the case of instantaneously applied supercritical normal field. In this case, at low field frequencies, the drop breakup into a large number of rotating small droplets, rings, and S-shaped droplets was observed (Figure 8a). Such manner of droplet disintegration has not been observed previously. At higher frequencies the formation of several curved stripes encompassing the individual tiny droplets occurs (see Figure 8b); this pattern is analogous to that observed by Elborai et al. (2005). Note that some of the ferrofluid microdrop instability patterns (spiral arms and large crests) demonstrate the similarity to the flow patterns of the thermal convection in a rotating fluid layer (Chandrasekhar 1981). The similar nature of these phenomena can be expected.

The nonmagnetic microdrop immersed in the ferrofluid also becomes disk-like shaped under the action of in-plane-rotating magnetic field (Dikansky and Zakinyan 2010). At comparatively low frequencies and high rotating field amplitudes, gradually rising normal magnetic field leads to the instability development in the form of curly crests appearing on the microdrop boundary (see Figures 9a, 9b and supplementary Online Resource 3). The development of crests leads to the microdrop breakup into a large number of smaller droplets. At high frequencies and low amplitudes the consecutive division of the initial microdrop into several smaller microdrops was observed under the action of normal field (Figure 9c). The phase diagram of corresponding instability patterns is presented in Figure 10. If the initial nonmagnetic microdrop is sufficiently large, the instability develops not only at the edge but also in the interior of disk-shaped microdrop. In this case the ferrofluid penetrates into the nonmagnetic droplet from above and from below forming the curly patterns and leading to the droplet breakup (see Figure 9d and supplementary Online Resource 3). If the normal magnetic field of comparatively high strength is applied instantaneously, the instability development in the internal area of the microdrop close to its center is more preferred (Figure 9e). Note that the behavior on nonmagnetic microdrop immersed in a ferrofluid layer under the simultaneous action of in-plane rotating magnetic field and normal magnetic field has not been analyzed previously. It should be also noted that in all the experiments described above we have not observed the hysteretic behavior of the studied systems.

**5 Discussion of the results**

The observed differences in the behavior of microdrops in comparison with the macroscopic cases should be associated with the distinguished properties of microscopic fluidic systems under study. Thus consider the relative ratio between the various forces involved in the process and having a profound impact in the experiments. The rotational Reynolds number is a ratio between the inertial and the viscous forces: $\mathrm{Re} = \rho \omega R^2 / \eta$, where $\eta$ is the dynamic viscosity of external fluid, $\rho$ the density of external fluid, $\omega$ the droplet angular rotation frequency. The rotational Weber number is a ratio between the inertial forces and the surface tension: $\text{We} = \rho\omega^2 R^3/\sigma$. For the microscopic cases considered here we have: Re ~ $10^{-4}$, We ~ $10^{-3}$. For the general macroscopic cases we have: Re ~ $10^3$, We ~ $10^4$. As it can be seen, in the microscopic case the viscous and surface tension effects dominate; contrary, in the macroscopic case the inertia influence is significant. The detailed theoretical analysis of the studied phenomena goes beyond the framework of the present experimental work; to this end we shall restrict our discussion to the consideration of a limited number of already existing theoretical models.

The instability conditions for a cylindrical ferrofluid drop and nonmagnetic drop in a horizontal layer have been derived from the second variation of the free energy functional by Blums et al. (1997). It was shown that the critical magnetic field decreases with the diameter increasing; and the nonmagnetic drop becomes unstable at higher fields than the ferrofluid drop, which correlates with our experiments. The effect of the drop breakup in a normal field can also be explained considering the total free energy of the drop: $F_1 = 2\pi R(\sigma h + \sigma' R) - \frac{\mu_0}{2}\chi H^2 V/(1+\chi D)$ (Elborai et al. 2005), where $\chi$ is the ferrofluid magnetic susceptibility, $H$ the magnetic field strength, $V$ the ferrofluid volume, $D = 1 - \frac{1}{2}h/\sqrt{R^2 + h^2/4}$ the demagnetizing factor of cylindrical droplet, $h$ the layer thickness, $R$ the droplet edge radius, $\sigma'$ the surface tension at fluid-solid interface. It can be easily demonstrated that at some value of the external field (the breakup field) the total free energy of a system of $N$ drops $F_N$ becomes lower than the free energy of a single drop with volume equal to the sum of volumes of $N$ drops. The expression for $F_N$ can be obtained from above expression for $F_1$ multiply it by $N$ and substituting $r = R/\sqrt{N}$ instead of $R$: $F_N = NF_1(R \to r)$. The breakup field can be obtained from the condition: $F_N = F_1$. It should be noted that we did not take into account the droplets interaction energy here. The total free energy of a system of interacting ferrofluid drops has been considered by Rosensweig (1985) and Elias et al (1997). It has been shown by Rosensweig (1985) and Elias et al (1997) that the droplets interaction leads to the increase of the critical magnetic field, which is in accordance with our experimental results.

The bending instability of a magnetic fluid stripe has been theoretically and numerically analyzed by Bacri et al. (1995). In the framework of linear theory the expression for the wavenumber of most unstable perturbations has been obtained. The corresponding calculations show that the instability wavelength decreases with the magnetic field and with the stripe thickness, which correlates with our experimental results on the nonmagnetic microstripes.

As it was mentioned above, the spiral arms and crests patterns of instability of the ferrofluid microdrop demonstrate the similarity to the flow patterns of the thermal convection in a rotating fluid layer. The spiral pattern of thermal convection in a rotating fluid layer is explained by the action of Coriolis force (Chandrasekhar 1981). Consider the volume element of ferrofluid droplet. The droplet angular rotation frequency can be estimated as $\omega \approx \frac{\mu_0}{2} \pi f \left(V_p/kT\right)\left(\eta_i/\eta_e\right) \chi H^2$ (Tsebers 1975), where $f$ is the magnetic field frequency, $V_p$ the magnetic particle volume. Elements of the ferrofluid drop repel each other due to the magnetic interaction in a normal magnetic field. The characteristic radial velocity of volume element associated with the magnetic force can be estimated from the dimensional analysis as $v^2 \sim \left(\mu_0 H^2/\rho\right)\left(\eta_i/\eta_e\right)(h/d)$, where $d$ is the droplet diameter. The trajectory of the volume element motion can be obtained from the set of equations: $\dot{x} = v\omega t \cos(\omega t)$, $\dot{y} = v\omega t \sin(\omega t)$. The corresponding calculations demonstrate that the volume element of ferrofluid droplet will move along the spiral trajectory. Note that only preliminary qualitative discussion of some obtained experimental results have been presented here; the detailed analysis of the studied phenomena is the goal of the future works.

## 6. Conclusion

In conclusion, we have explored the rich behavior of ferrofluid and nonmagnetic microdrops submitted to the constant and rotating magnetic fields. Enormously complex flow and instability patterns arise from a competition between the magnetic normal stress and the surface tension taken together with a competition between the tangential stress and the drop viscous torque in systems subjected to a geometric constraint. Here we summarize the obtained entirely new results and observations: the star-like patterns at the contact of the ferrofluid microdrop with the layer boundaries in a normal magnetic field; intensive breaking of ferrofluid microdrop through twisted trajectories in instantaneously applied normal magnetic field; simultaneous appearance of branching and bending instabilities of ferrofluid microstripes in a normal field; patterns of ridges, crests, rings, chaotically distorted droplet boundary and peculiarities of the droplet breakup under the combined action of stationary and rotating magnetic fields; all results of experiments with nonmagnetic droplet immersed in ferrofluid. The obtained results may be of interest in applications to the ferrofluid-based microfluidics and useful in the development of methods of material properties regulation by means of its structural microgeometry manipulation. The detailed theoretical analysis of investigated phenomena has not been performed in the present study due to their complexity; but we suggest that the experimental results presented here could stimulate further insights and theoretical studies.

**Supplementary materials**

The supplementary Online Resources contain the supporting video files.

**Video 1.** The instability and breakup of ferrofluid microdrop (left image) and nonmagnetic microdrop immersed in ferrofluid (right image) under the instantaneously applied perpendicular magnetic field of the strength 7.3 kA/m for ferrofluid microdrop and of the strength 9.5 kA/m for nonmagnetic microdrop.

**Video 2.** Shapes evolution of ferrofluid microdrops submitted to a counterclockwise magnetic field ($H_r$ = 5.5 kA/m) rotating in the plane of observation in the presence of a gradually rising perpendicular magnetic field ($H_n$). Left image: the rotating field frequency $f$ = 10 Hz. Right image: slow motion (speed reduced by a factor of two), $f$ = 70 Hz.

**Video 3.** Instability development of nonmagnetic microdrops immersed in a ferrofluid layer and submitted to a counterclockwise magnetic field rotating in the plane of observation in the presence of a gradually rising perpendicular magnetic field. Left image: $f$ = 70 Hz, $H_r$ = 5.5 kA/m, the perpendicular field ($H_n$) rises up to 4 kA/m. Right image: large nonmagnetic drop at $f$ = 10 Hz, $H_r$ = 2.3 kA/m, the perpendicular field ($H_n$) rises up to 2.1 kA/m.

**Acknowledgements**

This work was supported by the grant of the President of the Russian Federation (No. MK-5801.2015.1) and also by the Ministry of Education and Science of the Russian Federation in the framework of the base part of the governmental ordering for scientific research works (project 3.5822.2017).

**References**

Afkhami S, Tyler AJ, Renardy Y, Renardy M, Pierre TGSt, Woodward RC, Riffle JS (2010) Deformation of a hydrophobic ferrofluid droplet suspended in a viscous medium under uniform magnetic fields. J Fluid Mech 663:358–384

Bacri JC, Cebers A, Flament C, Lacis S, Melliti R, Perzynski R (1995) Fingering phenomena at bending instability of a magnetic fluid stripe. Prog Colloid Polym Sci 98:30–34

Bacri JC, Cebers A, Perzynski R (1994) Behavior of a magnetic fluid microdrop in a rotating magnetic field. Phys Rev Lett 72:2705–2708

Bacri JC, Levelut A, Perzynski R, Salin D (1988) Multiple scissions of ionic ferrofluid drops. Chem Eng Commun 67:205–216

Bacri JC, Salin D (1982) Study of the deformation of ferrofluid droplets in a magnetic field. J Phys Lett 43:L-179–L-184

Blums E, Cebers A, Maiorov MM (1997) Magnetic Fluids. de Gruyter

Chandrasekhar S (1981) Hydrodynamic and Hydromagnetic Stability. Dover

Chen CY, Tsai WK, Miranda JA (2008) Hybrid ferrohydrodynamic instability: Coexisting peak and labyrinthine patterns. Phys Rev E 77:056306

Chen CY, Wu WL, Miranda JA (2010) Magnetically induced spreading and pattern selection in thin ferrofluid drops. Phys Rev E 82:056321

Chen CY, Yang YS, Miranda JA (2009) Miscible ferrofluid patterns in a radial magnetic field. Phys Rev E 80:016314

Dickstein AJ, Erramilli S, Goldstein RE, Jackson DP, Langer SA (1993) Labyrinthine pattern formation in magnetic fluids. Science 261:1012–1015

Dikansky YuI, Zakinyan AR (2010) Dynamics of a nonmagnetic drop suspended in a magnetic fluid in a rotating magnetic field. Tech Phys 55:1082–1086

Drozdova VI, Skibin YuN, Chekanov VV (1981) Oscillation of a drop of magnetic liquid. Magnetohydrodynamics 17:320–324

Elborai S, Kim DK, He X, Lee SH, Rhodes S, Zahn M (2005) Self-forming, quasi-two-dimensional, magnetic-fluid patterns with applied in-plane-rotating and dc-axial magnetic fields. J Appl Phys 97:10Q303

Elias F, Flament C, Bacri JC, Neveu S (1997) Macro-organized patterns in ferrofluid layer: Experimental studies. J Phys I 7:711–728

Horng HE, Hong CY, Yang SY, Yang HC (2001) Novel properties and applications in magnetic fluids. J Phys Chem Solids 62:1749–1764

Jackson DP, Miranda JA (2007) Confined ferrofluid droplet in crossed magnetic fields. Eur Phys J E 23:389–396

Jackson DP, Goldstein RE, Cebers AO (1994) Hydrodynamics of fingering instabilities in dipolar fluids. Phys Rev E 50:298–307

Jamin T, Py C, Falcon E (2011) Instability of the origami of a ferrofluid drop in a magnetic field. Phys Rev Lett 107:204503

Janiaud E, Elias F, Bacri JC, Cabuil V, Perzynski R (2000) Spinning ferrofluid microscopic droplets. Magnetohydrodynamics 36:301–314

Lebedev AV, Engel A, Morozov KI, Bauke H (2003) Ferrofluid drops in rotating magnetic fields. New J Phys 5:57.1–57.20

Liu J, Tan SH, Yap YF, Ng MY, Nguyen NT (2011) Numerical and experimental investigations of the formation process of ferrofluid droplets. Microfluid Nanofluid 11:177–187.

Nguyen NT (2012) Micro-magnetofluidics: interactions between magnetism and fluid flow on the microscale. Microfluid Nanofluid 12:1–16.

Nguyen NT, Beyzavi A, Ng KM, Huang X (2007) Kinematics and deformation of ferrofluid droplets under magnetic actuation. Microfluid Nanofluid 3:571–579

Ody T, Panth M, Sommers AD, Eid KF (2016) Controlling the motion of ferrofluid droplets using surface tension gradients and magnetoviscous pinning. Langmuir 32:6967–6976

Rinaldi C, Chaves A, Elborai S, He X(T), Zahn M (2005) Magnetic fluid rheology and flows. Curr Opin Colloid Interface Sci 10:141–157

Rosensweig RE (1985) Ferrohydrodynamics. Cambridge University Press

Sandre O, Browaeys J, Perzynski R, Bacri JC, Cabuil V, Rosensweig RE (1999) Assembly of microscopic highly magnetic droplets: Magnetic alignment versus viscous drag. Phys Rev E 59:1736–1746

Tatulchenkov A, Cebers A (2005) Complex bubble dynamics in a vertical Hele-Shaw cell. Phys Fluids 17:107103

Timonen JVI, Latikka M, Leibler L, Ras RHA, Ikkala O (2013) Switchable static and dynamic self-assembly of magnetic droplets on superhydrophobic surfaces. Science 341:253–257

Torres-Diaz I, Rinaldi C (2014) Recent progress in ferrofluids research: novel applications of magnetically controllable and tunable fluids. Soft Matter 10:8584–8602

Tsebers AO (1975) Interfacial stresses in the hydrodynamics of liquids with internal rotation. Magnetohydrodynamics 11:63–66

Tsebers A, Blūms E (1988) Long-range magnetic forces in two-dimensional hydrodynamics of magnetic fluid pattern formation. Chem Eng Commun 67:69–88

Tsebers AO, Maiorov MM (1980) Magnetostatic instabilities in plane layers of magnetizable liquids. Magnetohydrodynamics 16:21–27

Wen CY, Lin JZ, Chen MY, Chen LQ, Liang TK (2011) Effects of sweep rates of external magnetic fields on the labyrinthine instabilities of miscible magnetic fluids. J Magn Magn Mater 323:1258–1262

Yu W, Bousmina M, Zhou C (2004) Determination of interfacial tension by the retraction method of highly deformed drop. Rheol Acta 43:342–349

Zakinyan A, Dikansky Y (2011) Drops deformation and magnetic permeability of a ferrofluid emulsion. Colloids Surf A 380:314–318

Zakinyan A, Tkacheva E, Dikansky Y (2012) Dynamics of a dielectric droplet suspended in a magnetic fluid in electric and magnetic fields. J Electrostat 70:225–232

Zakinyan AR, Dikansky YI (2017) Effect of microdrops deformation on electrical and rheological properties of magnetic fluid emulsion. J Magn Magn Mater 431:103–106

Zhu GP, Nguyen NT, Ramanujan RV, Huang XY (2011) Nonlinear deformation of a ferrofluid droplet in a uniform magnetic field. Langmuir 27:14834–14841

**Figures**

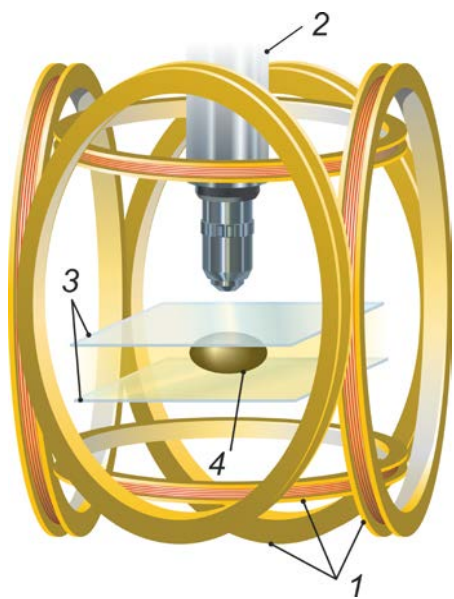

**Fig. 1** Layout of experimental setup: 1 – three pairs of perpendicular Helmholtz coils, 2 – microscope, 3 – two horizontal glass plates, 4 – microdrop.

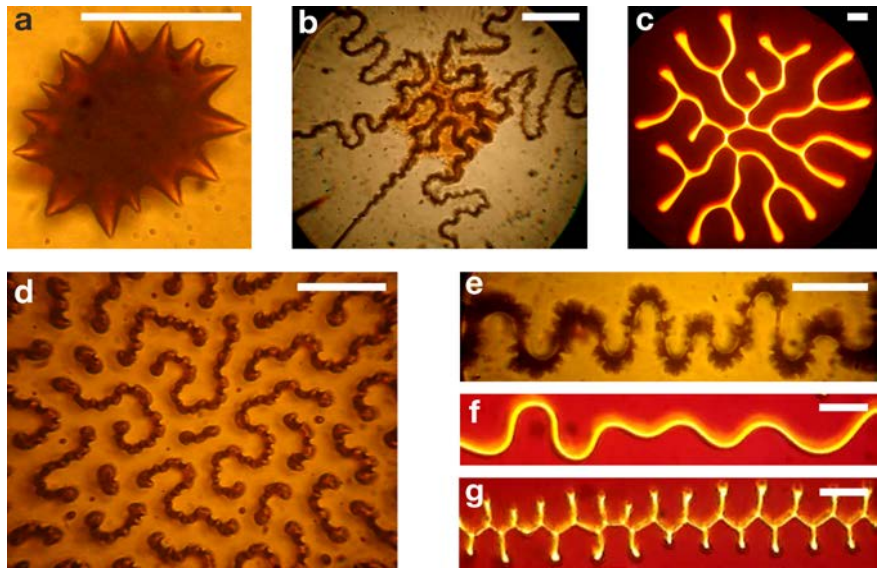

**Fig. 2** Shapes of microdrops submitted to a magnetic field oriented perpendicularly to the sample layer (plane of observation). (a) Stationary shape of a ferrofluid drop at the contact with upper plate (field strength 1.3 kA/m). (b) Snapshot of the intensively breaking ferrofluid drop at 1.8 s after instantaneous application of magnetic field (7.3 kA/m). (c) Stationary shape of a nonmagnetic drop immersed in a ferrofluid layer (field strength 2.6 kA/m). (d) Labyrinthine pattern formed by a large number of closely spaced ferrofluid drops (4.2 kA/m). Ferrofluid drops volume fraction is 40%. (e) Snapshot of the developing instability of a ferrofluid microscopic stripe at 3 s after instantaneous application of magnetic field (2.7 kA/m). (f) Bending instability of a nonmagnetic fluid microstripe in a ferrofluid layer (stationary shape at 7 kA/m applied instantaneously). (g) Fingering instability of a nonmagnetic fluid microstripe in a ferrofluid layer (stationary shape at 9 kA/m). All scale bars (white lines) are 50 μm

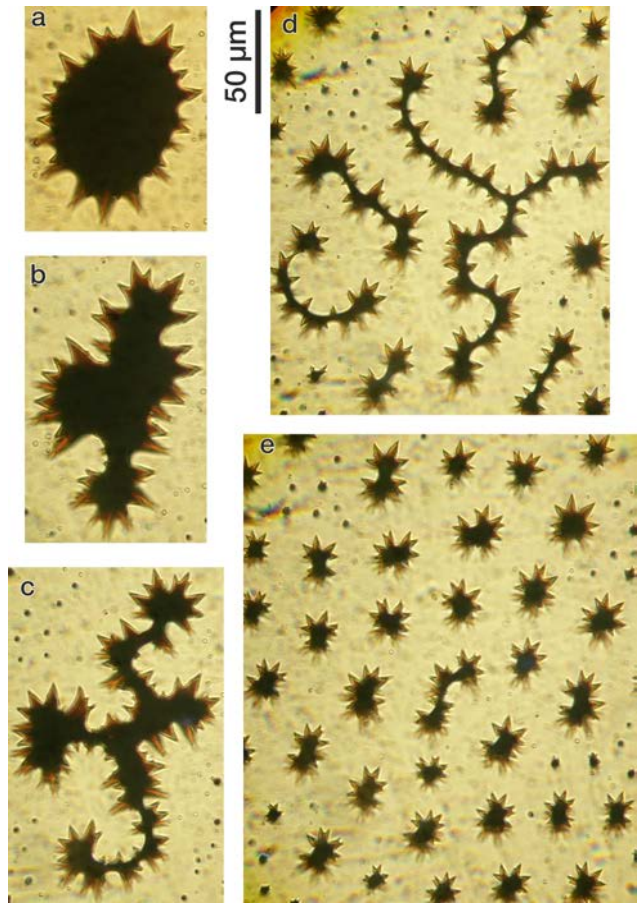

**Fig. 3** Instability and breakup of a ferrofluid microdrop confined in a flat layer under the action of gradually rising uniform perpendicular magnetic field. The stationary shapes at different values of the perpendicular magnetic field strength are presented: (a) 1 kA/m; (b) 1.5 kA/m; (c) 2 kA/m; (d) 2.5 kA/m; (e) 3 kA/m

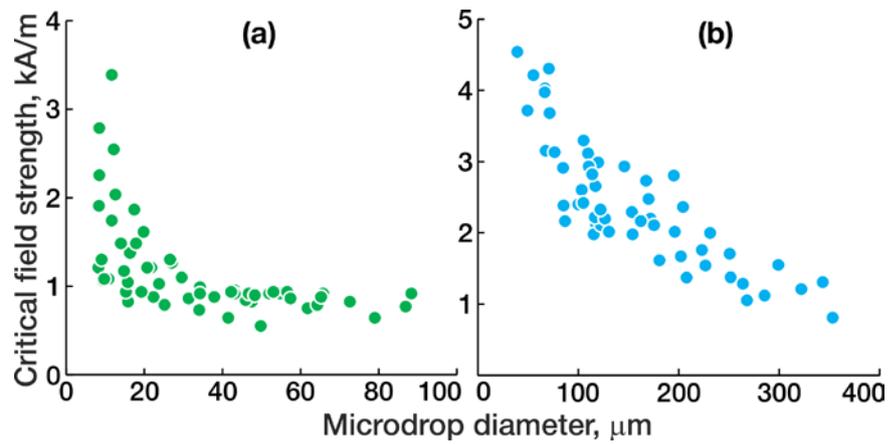

**Fig. 4** Experimental dependences of the critical perpendicular magnetic field strength at which the microdrop instability is taking place on the microdrop diameter visible from above in zero field. The dependences are obtained by visual observations in a gradually rising magnetic field. (a) The ferrofluid microdrops in a nonmagnetic liquid. (b) The nonmagnetic microdrops dispersed in ferrofluid

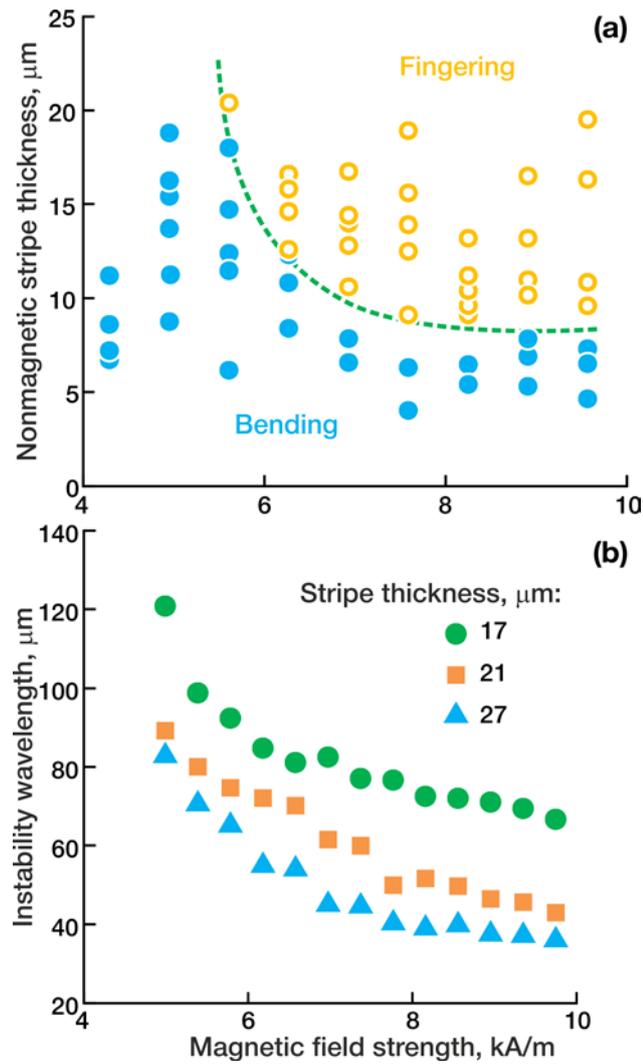

**Fig. 5** (a) Experimental phase diagram for the shape of the nonmagnetic microstripe in a ferrofluid layer under instantaneously applied perpendicular magnetic field. The lower-left region corresponds to the bending instability and the upper-right region corresponds to the fingering instability. (b) Experimental dependences of the nonmagnetic microstripe instability wavelength on the field strength at different values of the microstripe thickness. The absolute errors are of the order of the size of symbols representing experimental data

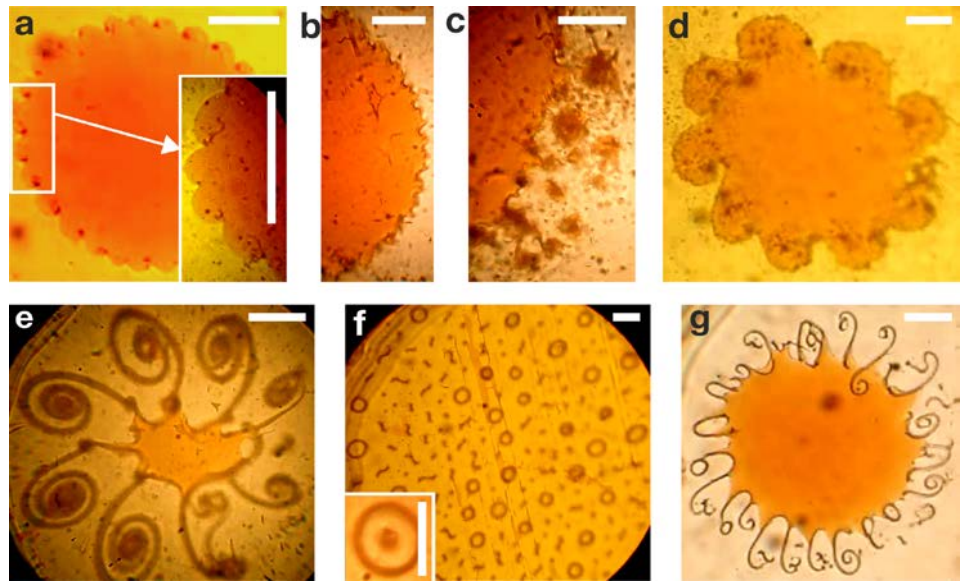

**Fig. 6** Shapes of ferrofluid microdrops submitted to a counterclockwise magnetic field rotating in the plane of observation in the presence of a gradually rising perpendicular magnetic field. (a) Liquid ridges at the rotating field frequency $f$ = 10 Hz, the rotating field amplitude $H_r$ = 3.6 kA/m and the perpendicular magnetic field strength $H_n$ = 1.4 kA/m. (b) Chaotically distorted boundary: $f$ = 10 Hz, $H_r$ = 3.6 kA/m, $H_n$ = 2.5 kA/m. (c) Breaking microdrop: $f$ = 10 Hz, $H_r$ = 3.6 kA/m, $H_n$ = 3 kA/m. (d) Large crests: $f$ = 30 Hz, $H_r$ = 5.5 kA/m, $H_n$ = 3.7 kA/m. (e) Snapshot of the microdrop with unstable spiral arms at 1.5 s after the instability development initiation: $f$ = 70 Hz, $H_r$ = 5.5 kA/m, $H_n$ = 4.5 kA/m. (f) The result of the microdrop breakup into several rotating rings and individual small droplets: $f$ = 70 Hz, $H_r$ = 5.5 kA/m, $H_n$ = 6 kA/m. Frequently appearing ring with a small droplet at the center is also shown. (g) Microdrop with stable spiral arms: $f$ = 150 Hz, $H_r$ = 7.7 kA/m, $H_n$ = 5 kA/m. All scale bars are 50 μm

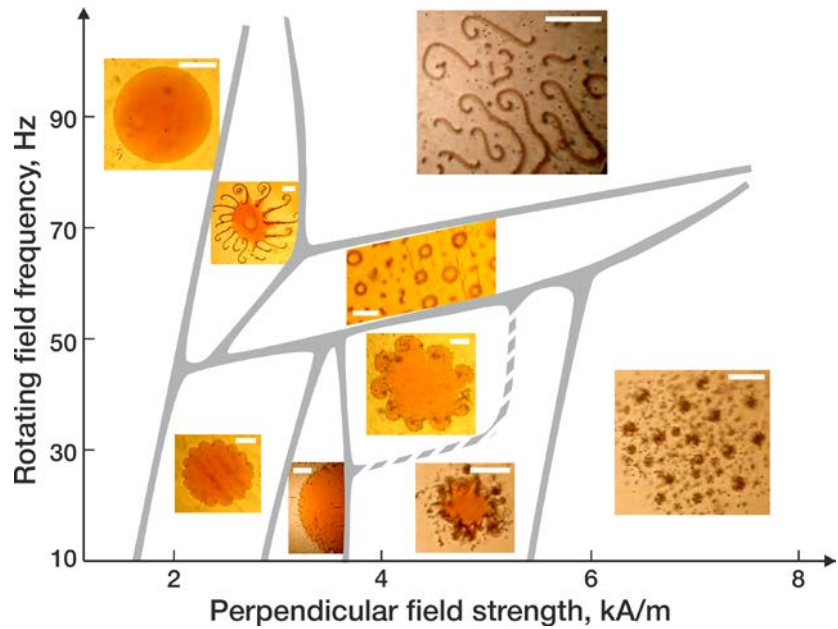

**Fig. 7** Qualitative phase diagram classifying the described shapes of ferrofluid microdrops submitted to the in-plane-rotating and gradually rising perpendicular magnetic fields at comparatively low rotating field amplitude. The large crests area outlined by dashed line appears at higher rotating field amplitudes. The areas of each microdrop configuration will shift up and to the right with increasing rotating field amplitude. All scale bars are 50 μm

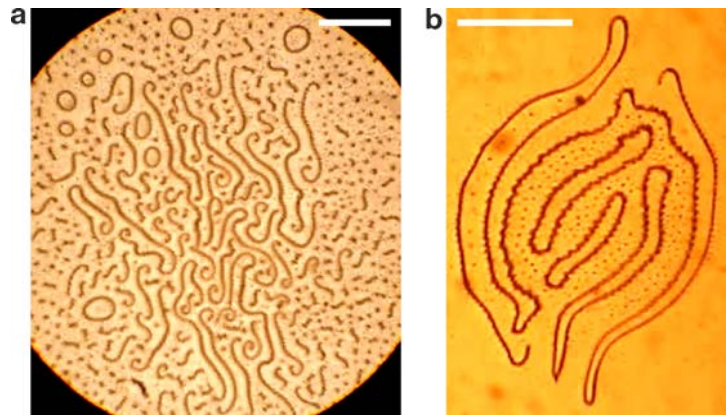

**Fig. 8** The breakup of ferrofluid microdrops submitted to a counterclockwise magnetic field rotating in the plane of observation after the instantaneous application of a perpendicular magnetic field. (a) Snapshot at 8.5 s after application of a perpendicular field: $f = 30$ Hz, $H_r = 3.6$ kA/m, $H_n = 7.5$ kA/m. (b) Snapshot at 1.3 s after application of a perpendicular field: $f = 150$ Hz, $H_r = 3.6$ kA/m, $H_n = 7.5$ kA/m. All scale bars are 100 μm

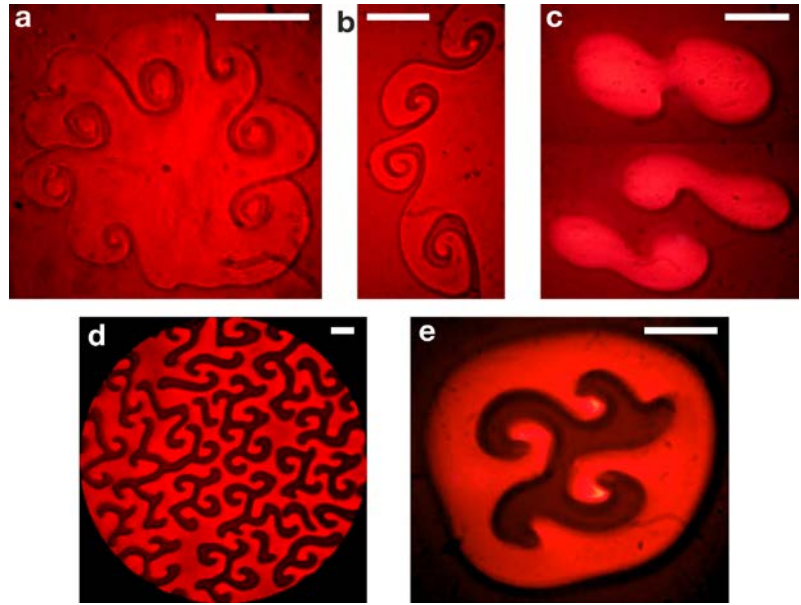

**Fig. 9** Instability patterns of nonmagnetic microdrops immersed in a ferrofluid layer and submitted to a counterclockwise magnetic field rotating in the plane of observation in the presence of a perpendicular magnetic field. (a) Snapshot of crests on the drop boundary at 3.8 s after the instability development initiation in a gradually rising perpendicular field: $f = 10$ Hz, $H_r = 3.6$ kA/m, $H_n = 2.1$ kA/m. (b) The same as in (a) at 1.5 s after the instability development initiation: $f = 70$ Hz, $H_r = 5.5$ kA/m, $H_n = 4.2$ kA/m. (c) The moments of consecutive divisions of the microdrop into smaller droplets under the action of normal field of the strength 5 kA/m (top image) and 6 kA/m (bottom image) ($f = 70$ Hz, $H_r = 3$ kA/m). (d) Snapshot of the instability pattern of large nonmagnetic drop at 3.6 s after the instability development initiation in a gradually rising perpendicular field: $f = 10$ Hz, $H_r = 2.3$ kA/m, $H_n = 2.3$ kA/m. (e) Snapshot of the instability pattern at 1 s after the instantaneous application of a perpendicular magnetic field: $f = 30$ Hz, $H_r = 5.5$ kA/m, $H_n = 7.5$ kA/m. All scale bars are 50 μm

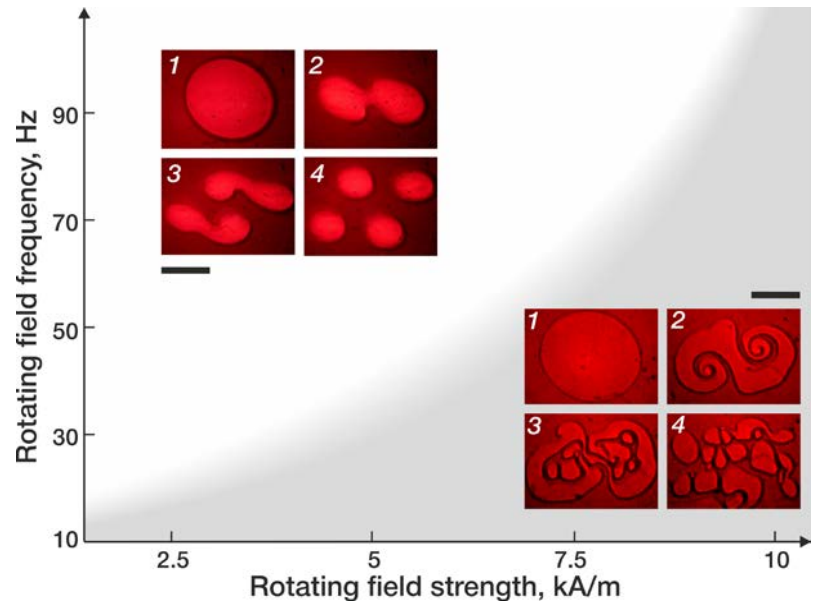

**Fig. 10** Qualitative phase diagram classifying the described shapes of nonmagnetic microdrops submitted to the in-plane-rotating and gradually rising perpendicular magnetic fields. The numbered images show the consecutive stages of the microdrop shape evolution with increasing perpendicular field. The lower-right region corresponds to the development of curly crests on the microdrop boundary and the upper-left region corresponds to the division of the microdrop into several smaller droplets. All scale bars are 50 μm